\documentclass[reprint,amsmath,amssymb,aps,twocolumn,superscriptaddress]{revtex4-1}
\usepackage{graphicx}% Include figure files
\usepackage{dcolumn}% Align table columns on decimal point
\usepackage{bm}% bold math
\usepackage{float}
\usepackage{amssymb}
\usepackage{color}
\usepackage{multirow}
\usepackage{stmaryrd}
\begin{document}

%\preprint{APS/123-QED}

\title{Anomalous Hall effect in distorted kagome magnets (Nd, Sm)Mn$_6$Sn$_6$}% Force line breaks with \\

\author{Wenlong Ma}
\affiliation{International Center for Quantum Materials, School of Physics, Peking University, Beijing 100871, China}
\author{Xitong Xu}
\affiliation{Anhui Key Laboratory of Condensed Matter Physics at Extreme Conditions, High Magnetic Field Laboratory, HFIPS, Anhui, Chinese Academy of Sciences, Hefei 230031, P. R. China}
\author{Zihe Wang}
\affiliation{International Center for Quantum Materials, School of Physics, Peking University, Beijing 100871, China}
\author{Huibin Zhou}
\affiliation{International Center for Quantum Materials, School of Physics, Peking University, Beijing 100871, China}
\author{Madalynn Marshall}
\affiliation{Department of Chemistry and Chemical Biology, Rutgers University, 123 Bevier Rd, Piscataway, NJ, 08854, USA}
\author{Zhe Qu}
\affiliation{Anhui Key Laboratory of Condensed Matter Physics at Extreme Conditions, High Magnetic Field Laboratory, HFIPS, Anhui, Chinese Academy of Sciences, Hefei 230031, P. R. China}
\affiliation{CAS Key Laboratory of Photovoltaic and Energy Conservation Materials, Hefei Institutes of Physical Sciences, Chinese Academy of Sciences, Hefei, Anhui 230031, China}
\author{Weiwei Xie}
\affiliation{Department of Chemistry and Chemical Biology, Rutgers University, 123 Bevier Rd, Piscataway, NJ, 08854, USA}
\author{Shuang Jia}%
 \email{gwljiashuang@pku.edu.cn}
\affiliation{International Center for Quantum Materials, School of Physics, Peking University, Beijing 100871, China}
\affiliation{Interdisciplinary Institute of Light-Element Quantum Materials and Research Center for Light-Element Advanced Materials, Peking University, Beijing 100871, China}
\affiliation{Collaborative Innovation Center of Quantum Matter, Beijing 100871, China}
\affiliation{CAS Center for Excellence in Topological Quantum Computation, University of Chinese Academy of Sciences, Beijing 100190, China}
\date{\today}% It is always \today, today,

\begin{abstract}

We report magnetic and electrical properties for single crystals of NdMn$_6$Sn$_6$ and SmMn$_6$Sn$_6$.
They crystallize into a structure which has distorted, Mn-based kagome lattices, compared to the pristine kagome lattices in heavy-rare-earth-bearing RMn$_6$Sn$_6$ compounds.
They are high-temperature ferromagnets of which the R moment is parallel with the Mn moment.
We observed a large intrinsic anomalous Hall effect (AHE) that is comparable to the ferrimagnetic, heavy-R siblings in a wide range of temperature.
We conclude that their intrinsic AHE is stemming from the Mn-based kagome lattice, just as in the heavy RMn$_6$Sn$_6$.

\end{abstract}

\pacs{Valid PACS appear here}% PACS, the Physics and Astronomy
\maketitle

%\tableofcontents

\section{\label{sec:level1}INTRODUCTION}
Research on the interplay of lattice geometry, magnetic structure, electron correlation and quantum topology is at the forefront of condensed matter physics~\cite{keimer2017physics,tokura2019magnetic,PhysRevLett.61.2015,chang2013experimental}.
Transition-metal-based kagome materials have attracted widespread attention because they often exhibit exotic quantum states including flat band~\cite{yin2019negative,kang2020dirac}, quantum spin liquid~\cite{PhysRevLett.98.107204,han2012fractionalized} and topological fermion~\cite{nakatsuji2015large,kuroda2017evidence,liu2018giant,ye2018massive,yin2018giant,liu2019magnetic,morali2019fermi}.
Made of corner sharing triangles, a magnetic kagome lattice possesses strong geometrical frustration, which may induce a novel quantum topological state~\cite{PhysRevB.62.R6065,ghimire2020topology}.
%Moreover, it hosts relativistic linear band crossings at the Brillouin zone corners~\cite{zhang2011quantum}, which characters the topological nontrivial electrical structure.
With the inclusion of spin-orbit coupling (SOC) and out-of-plane ferromagnetic (FM) ordering, a kagome lattice can effectively realize the spinless Haldane model generating Chern gapped Dirac fermion which renders a long-sought, high-temperature quantum anomalous Hall effect (QAHE) ~\cite{PhysRevLett.61.2015,PhysRevLett.106.236802,PhysRevLett.115.186802}.
However, pristine kagome lattices with strong out-of-plane magnetization are scarce in binary transition-metal compounds.

\begin{figure}[htbp]
	\begin{center}
		\includegraphics[clip, width=0.5\textwidth]{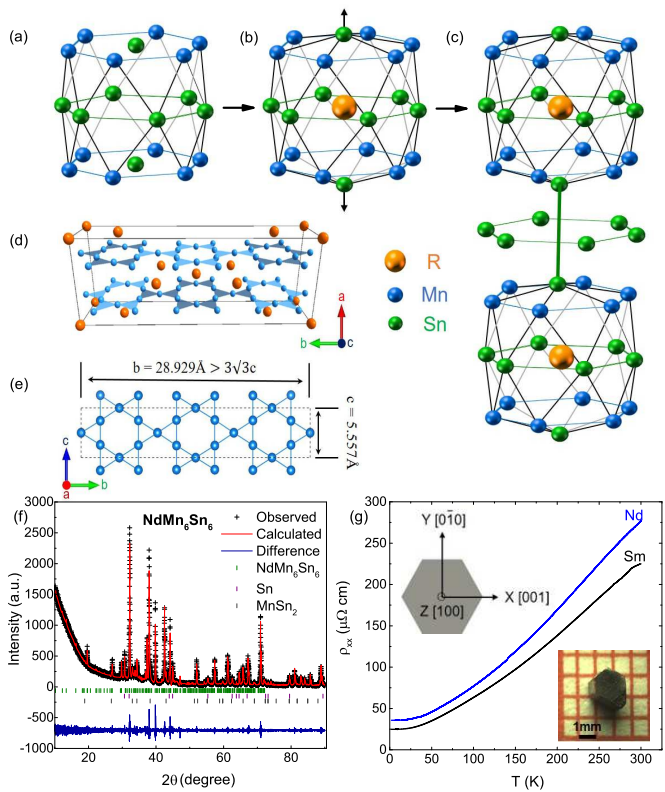}\\[1pt] % insert figure
		\caption{Crystal structure and resistivity of (Nd, Sm)Mn$_6$Sn$_6$. (a) to (c) describe the formation of RMn$_6$Sn$_6$ structure by stuffing an R ion into a CoSn-type structure. (a) An empty polyhedron in the CoSn-type structure. (b) R ion is stuffed into the polyhedron forming a pristine kagome lattice. (c) A layered RMn$_6$Sn$_6$ structure is formed~\cite{fredrickson2008origins}. (d) Distorted Mn-based kagome layers and R atoms in a unit cell of (Nd, Sm)Mn$_6$Sn$_6$. R atoms: yellow. Mn atoms: blue. (e) The distorted Mn-based kagome layer. (f) Refined powder XRD pattern of NdMn$_6$Sn$_6$. (g) Temperature dependence of in-plane, zero-field longitudinal resistivity. Top inset: Sketch of the crystallographic orientations. Bottom Inset: Photo of a single crystal of NdMn$_6$Sn$_6$ on 1 mm grid paper.}
		\label{fig:1}
	\end{center}
\end{figure}

A family of RMn$_6$Sn$_6$ (R = heavy rare earth element) compounds, crystallizing into a HfFe$_6$Ge$_6$-type structure ($P6/mmm$), host a pristine Mn-based kagome lattice, which generates various quantum magnetic properties~\cite{ma2020rare, yin2020discovery, ghimire2020competing, PhysRevB.101.100405,PhysRevB.103.014416,mielke2021intriguing}.
The R elements play an important role on the structural and magnetic properties of RMn$_6$Sn$_6$~\cite{VENTURINI199135,MALAMAN1999519,CLATTERBUCK199978}.
The HfFe$_6$Ge$_6$-type structure can be viewed as a R-stuffed CoSn-type structure (Fig. \ref{fig:1}(a)-(c)) \cite{venturini2006filling,fredrickson2008origins}.
Caged in the polyhedron made by Mn kagome and Sn honeycomb nets, the R elements push the Sn sites at the top and bottom of the void space away from the hexagonal center of the kagome net (Fig. \ref{fig:1}(b)).
This arrangement leads to an alternation of stuffed and empty cavities along the $c$ axis, which forms a layered structure (Fig. \ref{fig:1}(c)).
The pristine Mn-based kagome layer and the weak interlayer coupling in this layered structure are believed to facilitate the electron hopping in the kagome sites, which is crucial to realize the topological electron band~\cite{yin2020discovery}.

It is well-known that the spin parts of the $4f$ and $3d$ moments are prone to be coupled anti-parallel in many rare-earth transition metal intermetallics \cite{sinnema1984magnetic,brooks1991rare}.
Therefore the total moment of the Hund's ground state of the heavy R is coupled antiparallel to the transition metal moment whereas it is reverse for the light R. Just as observed in heavy RMn$_6$Sn$_6$ compounds, the magnetic R sublattice tends to develop an anti-parallel configuration with respect to the FM ordered Mn lattice
~\cite{MALAMAN1999519, CLATTERBUCK199978}.
Strong anti-parallel magnetic coupling between the anisotropic R moments and Mn moments leads to a ferrimagnetic (FIM) state in the RMn$_6$Sn$_6$ family (R = Gd - Er).
In particular, TbMn$_6$Sn$_6$, consisting of a high-temperature, out-of-plane FM ordered Mn-based kagome lattice, was discovered to be a near-ideal quantum-limit magnet with Chern-gapped, massive Dirac fermion ~\cite{yin2020discovery}.
Very recently we found the massive Dirac fermion generally exists in the FM Mn-based kagome lattices in RMn$_6$Sn$_6$ for R = Gd - Er ~\cite{ma2020rare}, whose Berry curvature field generates a large intrinsic anomalous Hall effect (AHE).
For comparison, the isostructural (Y, Lu)Mn$_6$Sn$_6$ with non-magnetic R atoms show a flat spiral antiferromagnetic (AFM) ordering of Mn moment~\cite{PhysRevB.103.094413,venturini1993magnetic}. A large topological Hall effect (THE) was observed in YMn$_6$Sn$_6$~\cite{PhysRevB.103.014416,ghimire2020competing}.

\begin{table*}
\caption{\label{tab:table1}Summary of the refined lattice parameters of the (Nd, Sm)Mn$_6$Sn$_6$ single crystals.}
\begin{ruledtabular}
\begin{tabular}{ccccccccc}
R & Space group & $Z$ & $a(\AA)$ & $b(\AA)$ & $c(\AA)$ & $V(\AA^3)$ \\
\hline
Nd & $Immm$ & 6 & 9.0616(3) & 28.9291(8) & 5.5575(2) & 1456.8650(5) \\
Sm (Phase1) & $Immm$ & 6 & 9.0649(5) & 28.9581(16) & 5.5870(3) & 1466.6002(8)  \\
Sm (Phase2) & $P6/mmm$ & 1 & 5.5487(1) & 5.5487(1) & 9.0438(2) & 241.1340(3)  \\
\end{tabular}
\end{ruledtabular}
\label{t1}
\end{table*}

\begin{figure*}[htbp]
	\begin{center}
		\includegraphics[clip, width=1\textwidth]{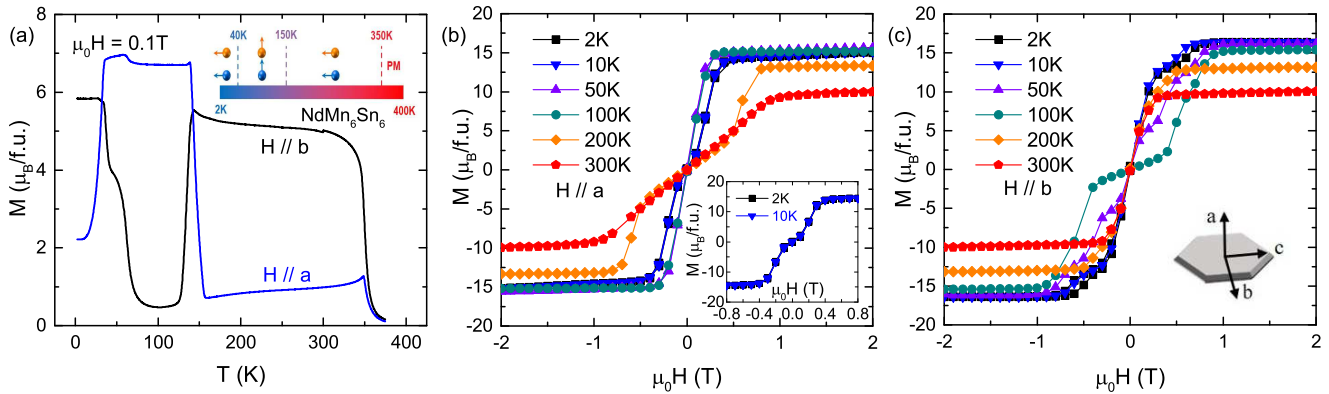}\\[1pt] % insert figure
		\caption{Magnetization ($M$) of NdMn$_6$Sn$_6$ . (a) Temperature dependence of $M$ with an external magnetic field ($\mu _0H=0.1 $~T) along crystallographic $a$ and $b$ directions. Inset: Magnetic structures in zero field at different temperatures. The blue and yellow spheres with arrows represent the Mn and Nd magnetic moments along different directions, respectively. (b) $M(H)$ curves at different temperatures when $H$ is applied along the $a$ axis. Inset shows the zoom-in $M(H)$ profiles at 2 and 10 K. (c) $M(H)$ curves at different temperatures when $H$ is applied along the $b$ axis. Inset shows the crystallographic orientations.}
		\label{fig:2}
	\end{center}
\end{figure*}

\begin{figure*}[htbp]
\begin{center}
\includegraphics[clip, width=1\textwidth]{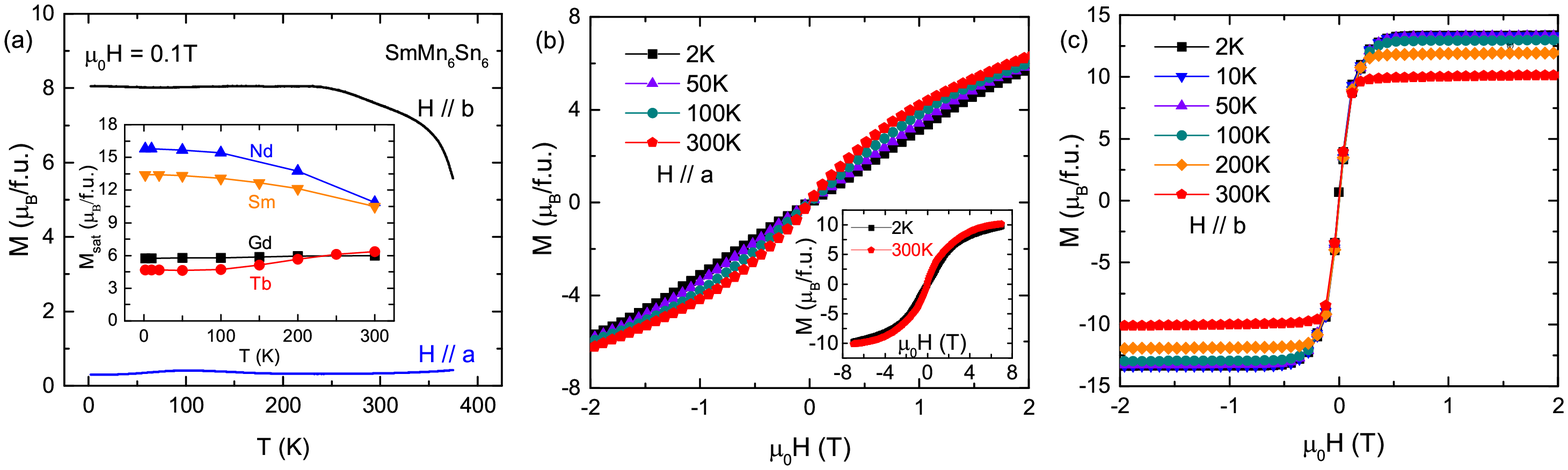}\\[1pt] % insert figure
		\caption{Magnetization ($M$) of SmMn$_6$Sn$_6$. (a) Temperature dependence of $M$ with an external magnetic field ($\mu _0H=0.1 $~T) along crystallographic $a$ and $b$ directions. Inset: Temperature dependence of the saturated magnetization ($M_{sat}$) for (Nd, Sm)Mn$_6$Sn$_6$ and (Gd, Tb)Mn$_6$Sn$_6$. (b) $M(H)$ curves at different temperatures when $H$ is applied along the $a$ axis. Inset shows the whole  profiles at 2 and 300 K. (c) $M(H)$ curves at different temperatures when $H$ is applied along the $b$ axis.}
\label{fig:3}
\end{center}
\end{figure*}

In this study we focus on (Nd, Sm)Mn$_6$Sn$_6$, two light rare earth siblings that have not been synthesized in a single-crystalline form until now, as far as we are aware.
Previous studies on its polycrystalline form showed several unique structural and magnetic properties unlike those in the heavy RMn$_6$Sn$_6$.
They crystallize into the HoFe$_6$Sn$_6$-type structure ($Immm$), another R-stuffed CoSn-type structure.
Unlike the HfFe$_6$Ge$_6$-type structure in which the R-stuffed polyhedron stacks along the $c$ axis in hexagonal cells, here the polyhedron forms triple rows along the $c$ axis and stacking along the $a$ axis in orthorhombic unit cells (Fig.~\ref{fig:1}(d)).
The triple rows of stuffed and empty cavities are alternatively arranged along the $b$ axis, leading to a larger unit cell which has a relation with the hexagonal cell as $a_{o}=c_{h}$, $b_{o}=3\sqrt{3}a_{h}$ and $c_{o}=b_{h}$ (Fig.~\ref{fig:1}(d))~\cite{venturini2006filling,fredrickson2008origins}.
%In HoFe$_6$Sn$_6$-type structure only Mn kagome lattice is preserved while the Sn atoms are no longer arranged in layer.
The structure distortion not only stretches the Mn kagome lattice along the $b$ axis, but also wrinkles it very slightly in the $a$ plane because the Mn atoms have four different crystallographic sites and coordinations now (Fig.~\ref{fig:1}(e)).
The magnetic coupling between the light R and Mn moment is parallel in (Nd, Sm)Mn$_6$Sn$_6$, leading to a collinear FM ordering above room temperature~\cite{malaman1997magnetic}.
Neutron diffraction on the poly-crystalline samples show SmMn$_6$Sn$_6$ is an easy-plane ferromagnet with the Curie temperature T$_{\mathrm{C}}$ = 405~K, while NdMn$_6$Sn$_6$ undergoes two spin-reorientation transitions below its T$_{\mathrm{C}}$ = 357~K~\cite{malaman1997magnetic}.

We investigate the magnetic anisotropy in single-crystalline (Nd, Sm)Mn$_6$Sn$_6$ and confirm the previously determined crystalline and magnetic structure in the polycrystals~\cite{weitzer1993structural,malaman1997magnetic}.
In general they are soft ferromagnets with a large saturated magnetic moment consisting of parallel R and Mn moments.
In particular, NdMn$_6$Sn$_6$ contains a distorted Mn-based kagome lattice with out-of-plane magnetization, similar to TbMn$_6$Sn$_6$.
(Nd, Sm)Mn$_6$Sn$_6$ possess a large intrinsic anomalous Hall conductivity (AHC) $\sigma_{AH}^{int}$ $\sim$ $100-200~\Omega^{-1}~\mathrm{cm}^{-1}$ below room temperature, which is comparable to the intrinsic AHC in FIM heavy RMn$_6$Sn$_6$~\cite{ma2020rare}.
Comparing the AHE in (Nd, Sm)Mn$_6$Sn$_6$ to that in heavy RMn$_6$Sn$_6$, we prove that the intrinsic AHE is stemming from the Mn lattice, while the scattering effect of R moments seems to be irrelevant.
%Although the crystal and magnetic structures are prominently changes, their Berry curvature field of the Mn kagome lattice seems to be preserved as that in the Chern gapped, massive Dirac fermion in heavy RMn$_6$Sn$_6$.

\begin{figure}[htbp]
\begin{center}
\includegraphics[clip, width=0.5\textwidth]{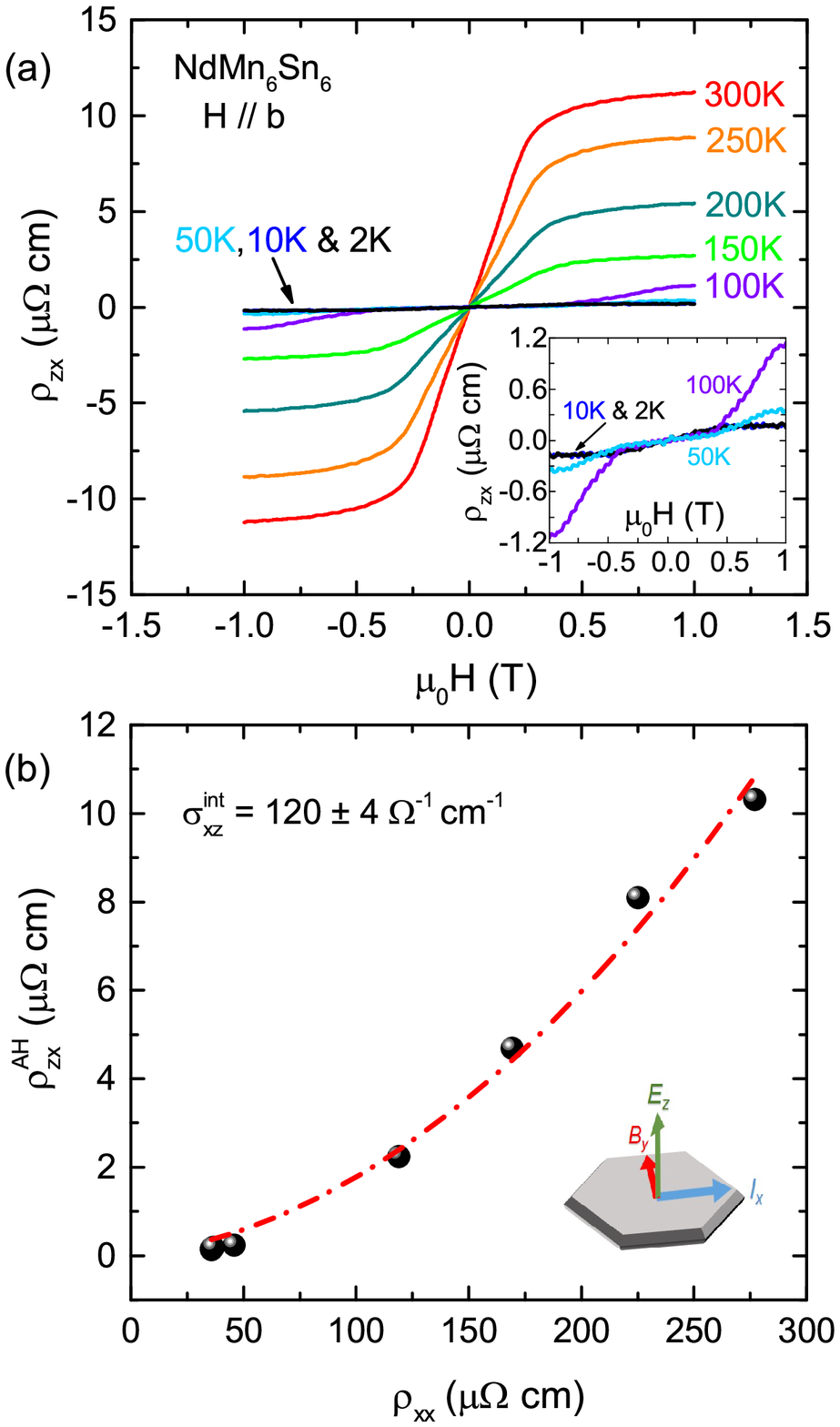}\\[1pt] % insert figure
\caption{AHE of NdMn$_6$Sn$_6$ when the external magnetic field is applied along the $b$ axis. (a) Field dependence of $\rho_{zx}$ at different temperatures. Inset: Zoom-in plot at low temperatures. (b) $\rho_{zx}^{AH}$ plotted against $\rho_{xx}$ from 2 K to 300 K. The red dashed line represents the polynomial fitting of the data points ($\rho_{AH} = \sigma^{int}\rho_{xx}^2 + \alpha^{skew}\rho_{xx}$), which gives the intrinsic anomalous Hall conductivity $\sigma_{xz}^{int}$. The bottom inset illustrates the geometry of the Hall measurement.}
\label{fig:4}
\end{center}
\end{figure}

\begin{figure}[htbp]
\begin{center}
\includegraphics[clip, width=0.5\textwidth]{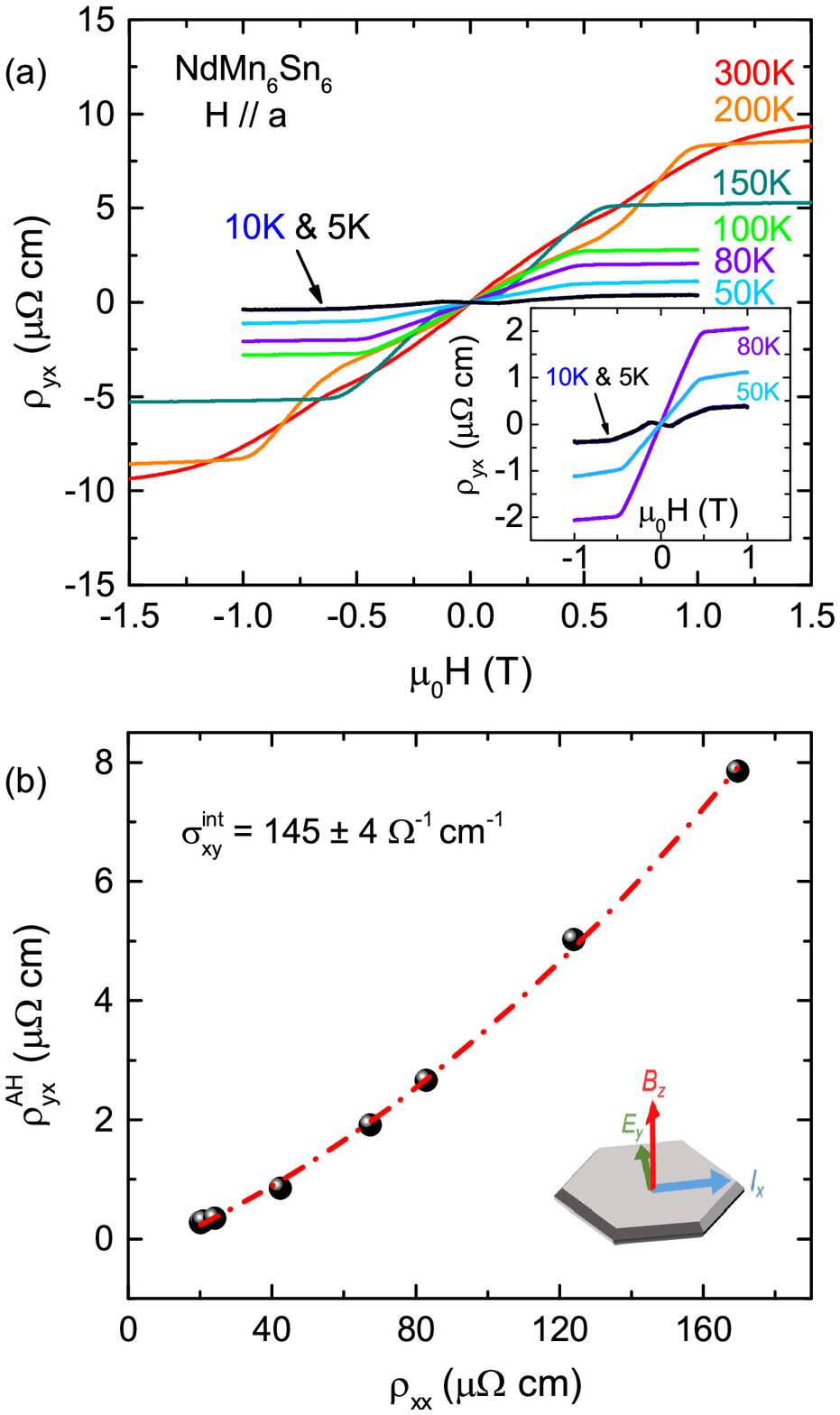}\\[1pt] % insert figure
\caption{AHE of NdMn$_6$Sn$_6$ when the external magnetic field is applied along the $a$ axis. (a) Field dependence of $\rho_{yx}$ at different temperatures. Inset: Zoom-in plot at at low temperatures. (b) $\rho_{yx}^{AH}$ plotted against $\rho_{xx}$ from 5 K to 300 K. $\sigma_{xy}^{int}$ is given by the polynomial fitting of the data points (red dashed line). The bottom inset illustrates the geometry of the Hall measurement.}

\label{fig:5}
\end{center}
\end{figure}

\begin{figure}[htbp]
\begin{center}
\includegraphics[clip, width=0.5\textwidth]{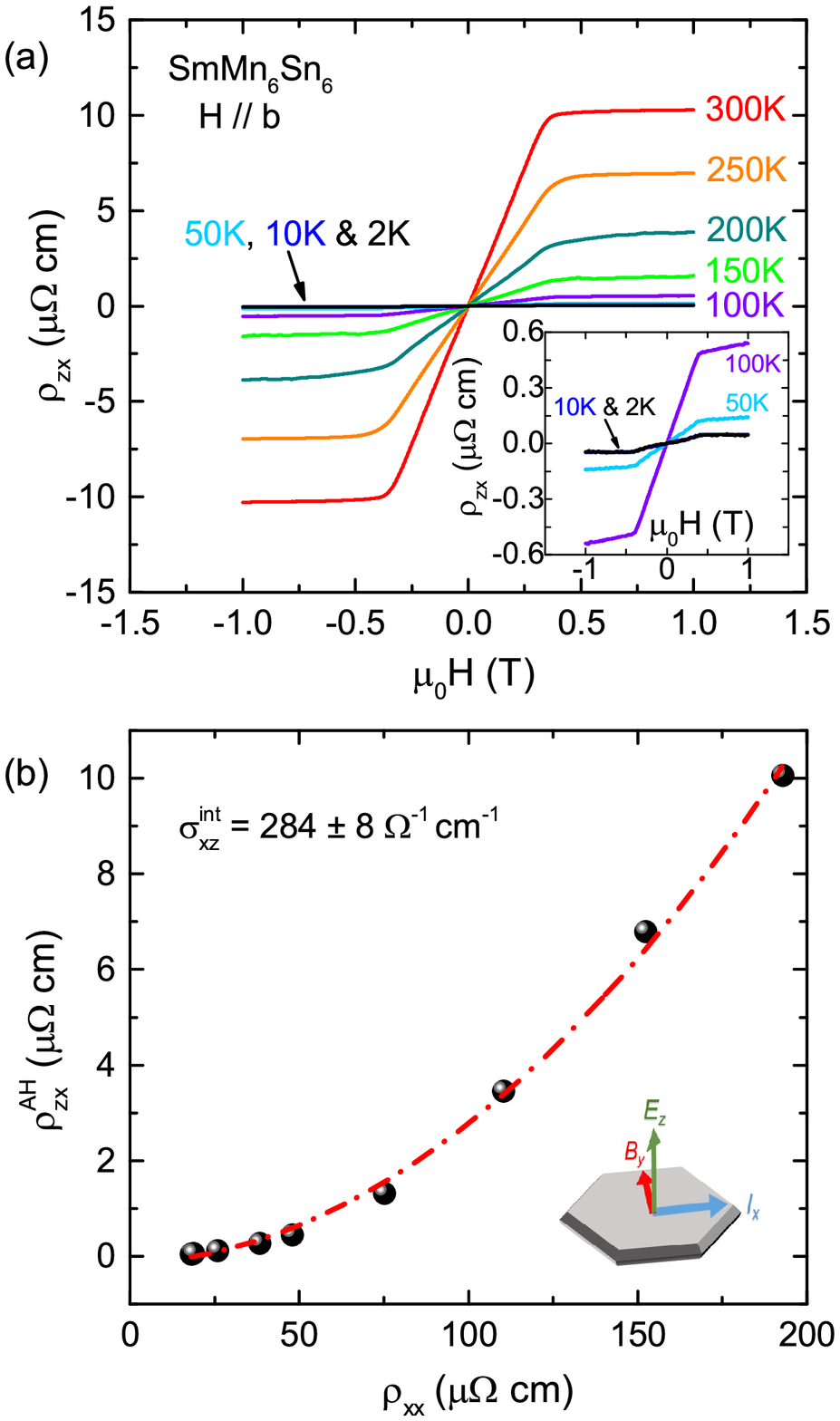}\\[1pt] % insert figure
\caption{AHE of SmMn$_6$Sn$_6$ when the external magnetic field is applied along the $b$ axis. (a) Field dependence of $\rho_{zx}$ at different temperatures. Inset: Zoom-in plot at low temperatures. (b) $\rho_{zx}^{AH}$ plotted against $\rho_{xx}$ from 2 K to 300 K. $\sigma_{xz}^{int}$ is given by the polynomial fitting of the data points (red dashed line). The bottom inset illustrates the geometry of the Hall measurement.}
\label{fig:6}
\end{center}
\end{figure}

\section{\label{sec:level2}EXPERIMENTAL RESULTS}

Single crystals of (Nd, Sm)Mn$_6$Sn$_6$ were synthesized by a tin-flux method~\cite{doi:10.1080/13642819208215073,CLATTERBUCK199978}.
The details of the crystal growth and experimental methods can be found in Sec. I of the Supplemental Material \cite{Supplemental_Material}.
Powder X-ray diffraction (PXRD) data collected on pulverized NdMn$_6$Sn$_6$ (Fig.~\ref{fig:1}(f)) was refined in the HoFe$_6$Sn$_6$-type structure plus a tiny amount of Sn and MnSn$_2$ impurities.
The PXRD pattern for SmMn$_6$Sn$_6$ shows two structures (HoFe$_6$Sn$_6$ and HfFe$_6$Ge$_6$-type) of crystals (Fig. S1 of the Supplemental Material \cite{Supplemental_Material}).
The result is consistent with the work on the polycrystal by Malaman \it et. al. \rm \cite{malaman1997magnetic} who reported that SmMn$_6$Sn$_6$ crystalizes into two structures at different synthesis temperatures.
We cannot distinguish the different structural crystals by appearance, however, the magnetization and transport measurements on several pieces show no explicit difference.
Therefore, we use the HoFe$_6$Sn$_6$-type structure to describe the sample below.

The typical sample morphology is a hexagonal prism (bottom inset of Fig.~\ref{fig:1}(g)) whose hexagonal surface is confirmed to be the (100) plane by a Laue diffraction.
The crystallographic orientations are shown in the upper inset of Fig.~\ref{fig:1}(g).
As shown in Table~\ref{t1} and Fig.~\ref{fig:1}(e), the ratio of lattice constants $b$ and $c$ for NdMn$_6$Sn$_6$ is slightly larger than $3\sqrt{3}$, reflecting the stretch of its kagome lattice.
It is noteworthy that the lattice parameters of SmMn$_6$Sn$_6$ are slightly larger than those of NdMn$_6$Sn$_6$.
Violation of the lanthanide contraction may indicate an unstable valence of Sm ions.

Temperature dependent, zero-field longitudinal resistivity $\rho_{xx}(T)$ of (Nd, Sm)Mn$_6$Sn$_6$ is presented in Fig.~\ref{fig:1}(g) when the current ($I$) is applied along the $c$ axis.
Their residual resistivity ratio ($RRR = \rho(300 \mathrm{K})/\rho(2 \mathrm{K})$) is around 10.
Because the resistivity $\rho_{zz}$ for $I\parallel a$ is close to  $\rho_{xx}$ at room temperature, we use an isotropic $\rho_{xx}$ to analysis the data below.

%indicating that the overall electronic structure is more 3D-like~\cite{sales2019quasi,asaba2020anomalous}.
A uniaxial anisotropy is observed in the magnetization curves $M(H)$ when $H$ is along three crystallographic orientations.
While the $M(H)$ curves for $H \parallel b$ and $c$ directions are identical (the later is shown in Fig. S3 of the Supplemental Material \cite{Supplemental_Material}), the anisotropic magnetic properties are demonstrated when $H$ is applied along the $a$ and $b$ directions (shown in Fig.~\ref{fig:2} and \ref{fig:3}).
On the whole our results are consistent with that of the magnetization and neutron scattering experiments on the polycrystals~\cite{malaman1997magnetic}.
Below we show the details of single crystals that were unresolved before.

According to the $M(T)$ curves, NdMn$_6$Sn$_6$ has an FM transition with a $T_{\mathrm{C}}$ of about 350 K, similar to 357 K observed for the polycrystal~\cite{malaman1997magnetic}  (Fig.~\ref{fig:2}(a)).
Below the $T_{\mathrm{C}}$, the $M(T)$ curves show two sharp spin-reorientation transitions in which the easy axis changes from in-plane to out-of-plane at about 150~K and then back to in-plane at about 40~K (Fig.~\ref{fig:2}(a) and inset).
These complicated spin-reorientation transitions were also detected in neutron scattering \cite{malaman1997magnetic}.
Yet we observed some peculiar features on the $M(T)$ curves between 60 and 40~K, which may be stemming from some unknown magnetic structure changes.

The spin-reorientation transitions are visible in a series of $M(H)$ curves when $H$ is along the $a$ and $b$ axes (Fig.~\ref{fig:2}(b) and (c)).
Overall the single crystal of NdMn$_6$Sn$_6$ is a soft magnet whose magnetization is saturated in a small field ($\mu_0H<$ 1 T) along all directions below room temperature.
The saturated magnetization ($M_{sat}$) is about 15 $\mu_B /\mathrm{f.u.}$ at 2~K, which is constituent of six Mn moments and one Nd moment in parallel (Fig.~\ref{fig:2}(b)).
$M_{sat}$ gradually decreases to 10 $\mu_B /\mathrm{f.u.}$ as the temperature increases to 300 K.
An external field less than 0.4 T along the hard direction drives the spin-reorientation when the temperature is below 40~K (inset of Fig.~\ref{fig:2}(b)).
At higher temperature the spin-reorientation requires a slightly larger external field, no more than 1~T at 300~K.

The magnetic structure of SmMn$_6$Sn$_6$ is rather simple as we observed an easy-plane FM state in the $M(T)$ curves for the whole temperature range (Fig.~\ref{fig:3}(a)).
Its $T_C$ was reported to be about 400~K~\cite{weitzer1993structural,malaman1997magnetic}, which is beyond our measurement range.
When the field is applied in-plane, the $M(H)$ curves show a typical soft FM profile and the $M_{sat}$ is about 13 $\mu_B /\mathrm{f.u.}$ at 2~K and gradually decreases to 10 $\mu_B /\mathrm{f.u.}$ at 300 K (Fig.~\ref{fig:3}(c)).
In an out-of-plane magnetic field, $M$ steadily increases and shows no trend of saturation at 7~T (Fig. ~\ref{fig:3}(b) and inset).

We compare the $M_{sat}$ of (Nd, Sm)Mn$_6$Sn$_6$ and (Gd, Tb)Mn$_6$Sn$_6$, which have easy-plane and easy-axis FIM states below their $T_{\mathrm{C}}$, respectively.
The $M_{sat}$ of (Gd, Tb)Mn$_6$Sn$_6$ gradually increases from 5.7 and 4.6 $\mu_B /\mathrm{f.u.}$ at 2 K to 6.0 and 6.4 $\mu_B /\mathrm{f.u.}$ at 300 K, respectively (inset of Fig.~\ref{fig:3}(a)).
If we assume the saturated moments of the R$^{3+}$ take the values of their Hund's rule ground states at 2 K (7 and 9 $\mu_B /\mathrm{f.u.}$ for Gd$^{3+}$ and Tb$^{3+}$, respectively), then we derive that the saturated moment of one Mn atom equals 2.1 and 2.3 $\mu_B$, respectively.
This result is comparable to the values in previous work~\cite{zhang2005unusual}.
Then we estimate the saturated moment of one Mn atom at 2 K for (Nd, Sm)Mn$_6$Sn$_6$ in the same way, which is about 2.0 and 2.1 $\mu_B$, respectively.
The opposite trend of the $M_{sat}$ with the temperature change for light and heavy R indicates that the $M_{sat}$ change in temperature is mainly due to the damping of the rare earth moment with increasing temperature.

Given the fast-saturating $M(H)$ curves, we speculate that the AHE can be easily detected in NdMn$_6$Sn$_6$ when a weak magnetic field is applied along any direction, but can only be detected in SmMn$_6$Sn$_6$ when a weak magnetic field is applied in-plane.
Unfortunately, we found that the crystals are prone to crack under a magnetic field higher than 1~T, which might be due to a large magnetostriction effect.
Please note we did not meet this difficulty when measuring the heavy R siblings.
Figure~\ref{fig:4}, \ref{fig:5} and \ref{fig:6} show the Hall resistivity, $\rho_{zx}$ when $H\parallel b$ and $\rho_{yx}$ when $H\parallel a$ for NdMn$_6$Sn$_6$ and $\rho_{zx}$ for SmMn$_6$Sn$_6$ at different temperatures, respectively.
To avoid the sample damage we restricted the external field to less than 1.5~T.

The $\rho_{zx}$ and $\rho_{yx}$ of NdMn$_6$Sn$_6$ follow the $M(H)$ isothermals in the whole temperature range, showing a large AHE above 100~K (Fig.~\ref{fig:4}(a) and \ref{fig:5}(a)).
The spin-reorientation transitions drive relevant changes in the $\rho_{zx}$ and $\rho_{yx}$ curves at various temperatures.
It is noteworthy that the $\rho_{yx}$ curves below 10~K are peculiar in weak field: they show a negative initial slope and a hump at about 0.1~T(inset of Fig.~\ref{fig:5}(a)), which is seemingly related to the spin-reorientation process at low temperatures (inset of Fig.~\ref{fig:2}(b)).
The reason for the negative initial slope in $\rho_{yx}$ below 10~K is still not clear.
In a stronger field $\rho_{yx}$ changes sign and is prone to saturate when $\mu_0H > 0.5$~T corresponding to the saturating external field of $M$.
The $\rho_{zx}$ of SmMn$_6$Sn$_6$ roughly resembles the $M(H)$ curves as well (Fig.~\ref{fig:6}(a)).
%except the wavy feature in low field below 250~K (Fig.~\ref{fig:6}(a)), which is absent in the $M(H)$ curves.
%This peculiar $\rho_{zx}$ profile might be due to some extra THE contribution when the Mn moments rotate in plane.
%Further study is needed to clarify it and below we focus on the AHE when $M$ is saturated.

\section{\label{sec:level3}ANALYSIS AND DISCUSSION}

Except for the peculiar feature mentioned above in NdMn$_6$Sn$_6$ below 10~K, the Hall resistivity can be related to the magnetization as described by the following equation~\cite{RevModPhys.82.1539}:
\begin{equation}
\rho_{yx} = \rho_{OH}(B)+\rho_{AH}(M)= R_0B+ R_S4\pi M
\label{eqn:1}
\end{equation}
where $B$ is the magnetic induction field, $\rho_{OH}$ and $\rho_{AH}$ are the ordinary and anomalous Hall resistivity, $R_0$ and $R_S$ are the ordinary and anomalous Hall coefficients, respectively.
%This relationship is impregnable and the AHE can be discerned even at the base temperature (insets of Fig.~\ref{fig:4}(a) to \ref{fig:6}(a)).

The AHE is known to have origins from intrinsic and extrinsic (skew-scattering and side-jump while the later is usually small) mechanisms~\cite{RevModPhys.82.1539}.
Since $M_{sat}$(T) changes small over the temperature range we investigate ($M_{sat}$(300K)/$M_{sat}$(2K) $\sim$ 0.7), we ignore the effect of the $M_{sat}$ change on the AHE~\cite{PhysRevLett.96.037204,PhysRevB.73.224435}.
In order to untangle the intrinsic term, we use the following equation~\cite{PhysRevLett.96.037204}:
\begin{equation}
\rho_{AH} = \sigma^{int}\rho_{xx}^2 + \alpha^{skew}\rho_{xx}
\label{eqn:2}
\end{equation}
where $\sigma^{int}$ is the intrinsic AHC and $\alpha^{skew}$ is the skew-scattering contribution parameter.
Using a standard polynomial fitting, we obtained $\sigma^{int}$ as about $130~\Omega^{-1}~\mathrm{cm}^{-1}$
for two directions in NdMn$_6$Sn$_6$ and about $284~\Omega^{-1}~\mathrm{cm}^{-1}$ in SmMn$_6$Sn$_6$ (Fig.~\ref{fig:4}(b) to \ref{fig:6}(b)).
The similar values of $\sigma^{int}$ can be obtained by using a modified Tian-Ye-Jin scaling method~\cite{PhysRevLett.103.087206}.
The values of $\alpha^{skew}$ for (Nd, Sm)Mn$_6$Sn$_6$ have the same magnitude of order as that for FIM heavy RMn$_6$Sn$_6$ ($\sim 5 \times 10^{-3}$)~\cite{ma2020rare}.
%This difference may reflect different spin orbital coupling energy and electron scattering time in different crystal structures~\cite{PhysRevB.77.165103}.
%As demonstrated in previous theoretical study, the major AHE mechanism is related to the SOC energy ($E_{SOC}$) and the electron scattering time ($\tau$) in ferromagnets~\cite{PhysRevB.77.165103}.
%The skew-scattering dominates the role in AHE mechanism when $\sigma_{xx}~\gg~\frac{ne^{2}\hbar}{m^{*}E_{SOC}}$ ($\hbar$ is the reduced Planck constant, $m^{*}$ is the effective electron mass, $n$ is the carrier density, and e is the electron charge), while the intrinsic contribution dominates when $\sigma_{xx}~\leq~\frac{ne^{2}\hbar}{m^{*}E_{SOC}}$.
%To evaluate the critical conductivity between two contributions, we calculate the skew-scattering AHC ($\sigma^{skew}$) by using $\sigma^{skew}$ = $\alpha^{skew}\sigma_{xx}$.
%Therefore, we can get the crossover of the AHE mechanism between the skew-scattering and intrinsic contributions with $\sigma_{xx}~\approx~1~\times~10^{4}~\Omega^{-1}~\mathrm{cm}^{-1}$ for $\sigma_{xy}$ of NdMn$_6$Sn$_6$. This value is smaller compared with that for other conventional ferromagnets with an order of $10^{5}~\sim~10^{6}~\Omega^{-1}~\mathrm{cm}^{-1}$~\cite{PhysRevB.77.165103}, which may be attributed to the different SOC energy and lower carrier density ($\sim~10^{21}~\mathrm{cm}^{-3}$) in NdMn$_6$Sn$_6$.

%HERE YOU ADD SOME SENTENCES TO DISCUSS WHAT IS THE ALPHA-SKEW AND COMPARE IT TO HEAVY 166 AND OTHER KAGOME MAGNETS.
%MAYBE STATE THAT THE EXTRINSIC IS SMALL COMPARING WITH INTRINSIC.

Comparing $\rho_{AH}$ and $\sigma_{xy}^{int}$ in NdMn$_6$Sn$_6$ to that in TbMn$_6$Sn$_6$, we find they have the same sign and are close in magnitude, albeit the $M_{sat}$ of NdMn$_6$Sn$_6$ in the FM state is several times larger than that of TbMn$_6$Sn$_6$ in FIM state.
This observation can be well understood if the AHE is mainly stemming from the Mn lattice while the scattering effect of R-moments is negligible.
The $\sigma_{xy}^{int}$ in NdMn$_6$Sn$_6$ corresponds to $(0.17~\pm~0.01)~e^2/h$ per kagome layer, close to the value of TbMn$_6$Sn$_6$ as well~\cite{yin2020discovery}.
Comparing their crystal structures, we notice the Sn and R atoms have been rearranged while the Mn kagome lattice only has a slight distortion.
We speculate that the crystal structure change does not significantly modify the Berry curvature field.
It needs further study to clarify whether and how the structural distortion impacts the massive Dirac fermion which was hosted in the Mn kagome lattice.

Another interesting observation is that SmMn$_6$Sn$_6$ seems to have a larger $\sigma_{xz}^{int}$ compared to that of GdMn$_6$Sn$_6$, which has an easy-plane FIM state~\cite{PhysRevB.101.174415}.
A plausible valance instability of the Sm ion may affect its electron structure.
%Because SmMn$_6$Sn$_6$ has large lattice parameter and a plausible valence instability of the Sm ion, the electronic structure probably differs from the other members.
The effect of the R valence instability on the topological electron properties of the RMn$_6$Sn$_6$, including SmMn$_6$Sn$_6$ and YbMn$_6$Sn$_6$, needs further elaboration.

%The value of total angular momentum $\mathbf{J}$ in the ground state is $\mathbf{L} \mp \mathbf{S}$ ($\mathbf{L}$ is the total orbital momentum and $\mathbf{S}$ is the total spin momentum), according as the subshell is less or more than half-full. $\mathbf{J} = \mathbf{L} - \mathbf{S}$ in the tripositive light rare earth ions since the 4f subshell is less than half-full, while $\mathbf{J} = \mathbf{L} + \mathbf{S}$ in the tripositive heavy rare earth ions.

\section{\label{sec:level4}CONCLUSION}

In summary, we investigate the magnetic and electrical properties of single-crystalline (Nd, Sm)Mn$_6$Sn$_6$ whose crystal structure features distorted Mn kagome lattices.
The compounds are soft, room-temperature ferromagnets with an easy-plane or easy-axis anisotropy with respect to the kagome plane.
They possess a large intrinsic AHE comparable to that in the FIM heavy RMn$_6$Sn$_6$.
Their intrinsic AHE is stemming from the Berry curvature field in Mn-based kagome lattice as well.

The kagome magnets are candidates for quantum materials, but the distorted kagome magnets were rarely studied.
Our observation demonstrates a large, Berry curvature field induced AHE in distorted kagome magnets.
Noticing that a large family of RT$_6$X$_6$ (T = transition metal, X = metalloid) compounds crystallize into various structures consisting of distorted and pristine T-based kagome lattices~\cite{schobinger1997ferrimagnetism,brabers1994magnetic,venturini1992crystallographic,rao1997mossbauer,mazet2001macroscopic,szytula2004crystal},
we will investigate their topological properties and the relationship with their lattice geometry and magnetic structures.

\section*{ACKNOWLEDGMENTS}
We would like to thank Prof. Yuan Li for using their wire cutting machine, Jia-Xin Yin for helpful discussions.
This work was supported by the National Natural Science Foundation of China No.U1832214, No.11774007, No.U2032213, No.11774352, the National Key R$\&$D Program of China (2018YFA0305601) and the strategic Priority Research Program of Chinese Academy of Sciences, Grant No. XDB28000000.
X. Xu acknowledges support from the China Postdoctoral Science Foundation No.2020M682056 and Anhui Postdoctoral Foundation No.2020B472.
X. Xu is also supported by CAS Key Laboratory of Photovoltaic and Energy Conservation Materials Foundation No.PECL2020ZZ007, Special Research Assistant, Chinese Academy of Sciences.
The work at Rutgers is supported by Beckman Young Investigator Award.

%\bigskip
%\bigskip
%\textbf{Acknowledgements}

\clearpage

\end{document}